\documentclass[11pt,a4paper]{article}

\newif\ifplainstyle
\plainstyletrue
\plainstylefalse

\newif\ifjhepstyle
\jhepstyletrue

\newif\ifprstyle
\prstyletrue
\prstylefalse

\ifjhepstyle
	\usepackage{jheppub}
	\usepackage{amsfonts}
	\usepackage{verbatim}
	\usepackage{float}
	\usepackage{color}
	\usepackage{array}
	\newcolumntype{C}[1]{>{\centering\arraybackslash$}p{#1}<{$}}
	\makeatletter
	\def\@fpheader{\phantom{Prepared for submission to JHEP}}
	\makeatother
\else	
 	\ifprstyle
		\usepackage{verbatim}
		\usepackage{amsmath,amsfonts,amssymb}
		\usepackage[colorlinks=true
                	,urlcolor=blue
                	,anchorcolor=blue
                	,citecolor=blue
                	,filecolor=blue
                	,linkcolor=blue
                	,menucolor=blue
                	]{hyperref}
	\else
            	\usepackage{verbatim}
            	\usepackage{cite}
            	\usepackage{setspace}
            	\usepackage[top=2.5cm, bottom=2.75cm, left=2.5cm, right=2.5cm]{geometry}
            	\usepackage{amsmath,amsfonts,amssymb}
            	\usepackage[colorlinks=true
            	,urlcolor=blue
            	,anchorcolor=blue
            	,citecolor=blue
            	,filecolor=blue
            	,linkcolor=blue
            	,menucolor=blue
            	,linktoc=page
            	]{hyperref}
            	\usepackage{float}
            	\restylefloat{table}
            	
            	\numberwithin{equation}{section}
	\fi
\fi

\allowdisplaybreaks

\newcommand{\ThisIsTheTitle}{Gauge and global symmetries of the candidate partially massless bimetric gravity}
\newcommand{\ThisIsAuthorOne}{Luis Apolo}
\newcommand{\ThisIsEmailOne}{luis.apolo@fysik.su.se}
\newcommand{\ThisIsAuthorTwo}{S.~F. Hassan}
\newcommand{\ThisIsEmailTwo}{fawad@fysik.su.se}
\newcommand{\ThisIsAuthorThree}{Anders Lundkvist}
\newcommand{\ThisIsEmailThree}{anders.lundkvist@fysik.su.se}
\newcommand{\ThisIsTheAffiliation}{Department of Physics \& The Oskar Klein Centre, \\
Stockholm University, AlbaNova University Centre, SE-106 91 Stockholm, Sweden}
\newcommand{\TheseAreTheKeywords}{}

\newcommand{\ThisIsTheAbstract}{In this paper we investigate a particular ghost-free bimetric theory that exhibits the partially massless (PM) symmetry at quadratic order. At this order the global $SO(1,4)$ symmetry of the theory is enhanced to $SO(1,5)$. We show that this global symmetry becomes inconsistent at cubic order, in agreement with a previous calculation. Furthermore, we find that the PM symmetry of this theory cannot be extended beyond cubic order in the PM field. More importantly, it is shown that the PM symmetry cannot be extended to quartic order in any theory with one massless and one massive spin-2 fields.}

\ifjhepstyle
\title{\ThisIsTheTitle}

\author{\ThisIsAuthorOne,}
\author{\ThisIsAuthorTwo}
\author{and \ThisIsAuthorThree}

\affiliation{\ThisIsTheAffiliation}

\emailAdd{\ThisIsEmailOne}
\emailAdd{\ThisIsEmailTwo}
\emailAdd{\ThisIsEmailThree}

\abstract{\ThisIsTheAbstract} 

\fi

\begin{document}

\ifjhepstyle
\maketitle
\flushbottom
\fi

\long\def\symfootnote[#1]#2{\begingroup%
\def\thefootnote{\fnsymbol{footnote}}\footnote[#1]{#2}\endgroup} 

\def\rednote#1{{\color{red} #1}}
\def\bluenote#1{{\color{blue} #1}}

\def\({\left (}
\def\){\right )}
\def\lb{\left [}
\def\rb{\right ]}
\def\lB{\left \{}
\def\rB{\right \}}

\def\Int#1#2{\int \textrm{d}^{#1} x \sqrt{|#2|}}
\def\Bra#1{\left\langle#1\right|} 
\def\Ket#1{\left|#1\right\rangle}
\def\BraKet#1#2{\left\langle#1|#2\right\rangle} 
\def\Vev#1{\left\langle#1\right\rangle}
\def\Vevm#1{\left\langle \Phi |#1| \Phi \right\rangle}\def\bbox{\bar{\Box}}
\def\til#1{\tilde{#1}}
\def\wtil#1{\widetilde{#1}}
\def\ph#1{\phantom{#1}}

\def\ra{\rightarrow}
\def\la{\leftarrow}
\def\lra{\leftrightarrow}
\def\p{\partial}
\def\diff{\mathrm{d}}

\def\sinh{\mathrm{sinh}}
\def\cosh{\mathrm{cosh}}
\def\tanh{\mathrm{tanh}}
\def\coth{\mathrm{coth}}
\def\sech{\mathrm{sech}}
\def\csch{\mathrm{csch}}

\def\a{\alpha}
\def\b{\beta}
\def\g{\gamma}
\def\d{\delta}
\def\e{\epsilon}
\def\ve{\varepsilon}
\def\k{\kappa}
\def\l{\lambda}
\def\n{\nabla}
\def\om{\omega}
\def\s{\sigma}
\def\t{\theta}
\def\z{\zeta}
\def\vp{\varphi}

\def\ss{\Sigma}
\def\dd{\Delta}
\def\gg{\Gamma}
\def\ll{\Lambda}
\def\tt{\Theta}

\def\A{{\cal A}}
\def\B{{\cal B}}
\def\cE{{\cal E}}
\def\D{{\cal D}}
\def\F{{\cal F}}
\def\H{{\cal H}}
\def\I{{\cal I}}
\def\J{{\cal J}}
\def\K{{\cal K}}
\def\L{{\cal L}}
\def\O{{\cal O}}
\def\P{{\cal P}}
\def\cS{{\cal S}}
\def\W{{\cal W}}
\def\X{{\cal X}}
\def\Z{{\cal Z}}

\def\we{\wedge}

\def\tilw{\tilde{w}}
\def\tile{\tilde{e}}

\def\zz{\bar z}
\def\xx{\bar x}
\def\xp{x^{+}}
\def\xm{x^{-}}

\def\VirU1{\mathrm{Vir}\otimes\hat{\mathrm{U}}(1)}
\def\VirSL2R{\mathrm{Vir}\otimes\widehat{\mathrm{SL}}(2,\mathbb{R})}
\def\U1{\hat{\mathrm{U}}(1)}
\def\SL2R{\widehat{\mathrm{SL}}(2,\mathbb{R})}
\def\sl2r{\mathrm{SL}(2,\mathbb{R})}
\def\by{\mathrm{BY}}

\def\RR{\mathbb{R}}

\def\tr{\mathrm{tr}}
\def\bnabla{\overline{\nabla}}

\def\sint{\int_{\ss}}
\def\dsint{\int_{\p\ss}}
\def\hint{\int_{H}}

\newcommand{\eq}[1]{\begin{align}#1\end{align}}
\newcommand{\eqst}[1]{\begin{align*}#1\end{align*}}
\newcommand{\eqsp}[1]{\begin{equation}\begin{split}#1\end{split}\end{equation}}

\newcommand{\absq}[1]{{\sqrt{-#1}}}



\ifprstyle
\title{\ThisIsTheTitle}

\author{\ThisIsAuthorOne}
\email{\ThisIsEmailOne}

\author{and \ThisIsAuthorTwo}
\email{\ThisIsEmailTwo}

\affiliation{\ThisIsTheAffiliation}


\begin{abstract}
\ThisIsTheAbstract
\end{abstract}


\maketitle

\fi



\ifplainstyle
\begin{titlepage}
\begin{center}

\ph{.}

\vskip 4 cm

{\Large \bf \ThisIsTheTitle}

\vskip 1 cm

\renewcommand*{\thefootnote}{\fnsymbol{footnote}}

{{\ThisIsAuthorOne}\footnote{\ThisIsEmailOne}, {\ThisIsAuthorTwo}\footnote{\ThisIsEmailTwo} and {\ThisIsAuthorThree}\footnote{\ThisIsEmailThree}}

\renewcommand*{\thefootnote}{\arabic{footnote}}

\setcounter{footnote}{0}

\vskip .75 cm

{\em \ThisIsTheAffiliation}

\end{center}

\vskip 1.25 cm

\begin{abstract}
\noindent \ThisIsTheAbstract
\end{abstract}

\end{titlepage}

\newpage

\fi

\ifplainstyle
\tableofcontents
\noindent\hrulefill
\bigskip
\fi


\section{Introduction}
\label{se:intro}

Non-linear partially massless gravity is an elusive theory closely related to massive~\cite{deRham:2010ik,deRham:2010kj,Hassan:2011vm} and bimetric gravity~\cite{Hassan:2011zd,Hassan:2011ea} but propagating fewer degrees of freedom.\footnote{For reviews on massive and bimetric gravity see~\cite{Hinterbichler:2011tt,deRham:2014zqa,Schmidt-May:2015vnx}.} At quadratic order in the action this is achieved via an additional gauge symmetry that removes the scalar mode of the massive spin-2 field described in both of these theories~\cite{Deser:1983mm,Deser:2001pe,Deser:2001us,Deser:2001wx,Deser:2001xr,Zinoviev:2001dt,Deser:2004ji}. A particularly appealing feature of this theory is that it requires a positive cosmological constant that is proportional to the squared mass of the spin-2 field. Since all corrections to the latter must be proportional to the mass itself, the cosmological constant can be naturally small~\cite{deRham:2013wv}.

The existence of this theory is in doubt, however, as several no-go results prevent extensions of the partially massless (PM) gauge symmetry beyond quadratic order~\cite{Zinoviev:2006im,Deser:2013uy,deRham:2013wv,Garcia-Saenz:2014cwa,Garcia-Saenz:2015mqi}. In particular, in~\cite{Zinoviev:2006im,deRham:2013wv} it was shown that massive gravity suffers from an obstruction at fourth order in the action that prevents the extension of the PM symmetry beyond cubic order. Different approaches where one considers either one or a multiplet of partially massless fields yield similar no-go results~\cite{Deser:2013uy,Garcia-Saenz:2014cwa,Garcia-Saenz:2015mqi}. 

An alternative proposal is to consider partially massless gravity as a
theory that requires an additional \emph{massless} spin-2 field,
i.e.~as a bimetric theory with a particular set of coupling
constants~\cite{Hassan:2012gz,Hassan:2012rq,Hassan:2013pca,Hassan:2015tba}. There
the PM symmetry is seen to exist up to sixth order in a derivative
expansion of the equations of
motion~\cite{Hassan:2015tba}. Furthermore, in a de Sitter background, a global version of the local PM symmetry is found to \emph{all orders} in the fields~\cite{Hassan:2012gz}. If the massless spin-2 field transforms non-trivially under the PM symmetry it may then be possible to overcome the results found in massive gravity.\footnote{This is somewhat reminiscent of higher spin theories where the massless spin-2 field necessarily transforms under the higher spin gauge symmetry.} Indeed, variation of the action under the PM transformation yields
  \eq{
  \d S \propto \int d^4x \(  \gg^{\mu\nu} \d \vp_{\mu\nu} + \Pi^{\mu\nu} \d h_{\mu\nu}  \), \label{se1:variation}
  }
where $\gg^{\mu\nu}$ and $\Pi^{\mu\nu}$ denote the variations with respect to the massive $\vp_{\mu\nu}$ and massless $h_{\mu\nu}$ fields. In principle the second term in eq.~\eqref{se1:variation} can counter the obstruction found in massive gravity at fourth order.

A similar setup has recently been considered in~\cite{Joung:2014aba}
which studies the global symmetries of the theory of a massless spin-2 field and a partially massless graviton order by order in the fields. Therein it is shown that, while the theory admits a global $SO(1,5)$ symmetry to lowest order in the fields, the global symmetry algebra fails to close at cubic order. Crucially, this inconsistency of the global symmetry algebra implies the inconsistency of the local PM symmetry beyond cubic order. The analysis of~\cite{Joung:2014aba} does not directly apply to the candidate PM bimetric theory, however. The reasons are twofold: first, the cubic order Lagrangian is not derived directly from the bimetric theory; and second, the set of parameters considered in~\cite{Joung:2014aba} actually make the cubic order Lagrangian of the bimetric theory vanish.

In this paper we analyze the gauge and global symmetries of the candidate PM bimetric theory beyond quadratic order in the fields. We find that at cubic order the bimetric action reduces to that considered in~\cite{Joung:2014aba} after appropriate field redefinitions and choice of parameters. The analysis of global symmetries then reveals that the $SO(1,5)$ symmetry algebra does not close in agreement with the general results of~\cite{Joung:2014aba}. The only way to satisfy the closure condition is to consider parameters in the theory where the distinction between the massless and the would-be partially massless fields becomes degenerate and the expansion of the action breaks down. Interestingly, this is precisely the set of parameters for which the equations of motion of the bimetric theory reproduce the equations of motion of conformal gravity at lowest order in a derivative expansion~\cite{Hassan:2013pca,Hassan:2015tba}. Since bimetric gravity is an otherwise consistent theory to all orders, we conclude that the enhanced $SO(1,5)$ symmetry seen at quadratic order is accidental. Hence the candidate PM theory admits only the standard $SO(1,4)$ global symmetries associated with the background de Sitter spacetime.

Next we consider the PM gauge symmetry of the bimetric theory and attempt to extend it beyond quadratic order in the massive spin-2 field. The non-closure of the global symmetry algebra would imply the absence of local PM symmetry beyond cubic order in the PM field~\cite{Joung:2014aba}. However, this analysis is not valid for all the possible parameters of the bimetric theory, particularly for the choice of parameters that make all odd powers of the PM field vanish. Ideally, since the bimetric action is known to all orders, one could search for the additional constraint that is responsible for removing one of the degrees of freedom from the spectrum. This is a technically challenging task due to the square root structure characterizing the potential of massive and bimetric gravity. Furthermore, as recalled above, an order-by-order construction of the PM gauge symmetry already fails at fourth order in massive gravity. Hence it proves rewarding to check first whether the bimetric theory is able to overcome this ``fourth order wall''. A hint that this may be possible is given by eq.~\eqref{se1:variation} provided that the massless spin two field transforms non-trivially under the PM symmetry. Although the massless field does transform non-trivially, we find that the local PM transformations cannot be extended beyond cubic order in the bimetric theory for any choice of parameters, in agreement with ref.~\cite{Joung:2014aba}. In fact, there is no quartic order action that would allow us to do so. Among all the possible quartic order actions, the one from bimetric gravity comes the closest to realizing the partially massless symmetry at non-linear order.

The paper is organized as follows. In Section~\ref{se:bimetric} we present the candidate PM bimetric theory and expand the action up to cubic order in the massive field. In Section~\ref{se3:pmsymmetry} we find the non-linear transformations of the massless and massive fields that leave the action invariant up to cubic order. In Section~\ref{se:global} we consider the algebra of global symmetries and show that the enhanced global symmetry algebra is inconsistent beyond quadratic order. Finally in Section~\ref{se:fourthorder} we consider the action to fourth order and show that no non-linear extension of the PM symmetry can keep the action invariant, despite the additional contributions from the massless spin-2 field. Furthermore, we find that no quartic action can render the action invariant under the PM symmetry. We end with our conclusions in Section~\ref{se:conclusions}. The explicit expression for the fourth order action is given in Appendix~\ref{ap:fourthorder}.


\section{Bimetric gravity and partial masslessness}
\label{se:bimetric}

In this section we present the quadratic theory of a partially massless field and introduce the candidate partially massless bimetric theory of ref.~\cite{Hassan:2012gz}. We then expand the action of the latter up to cubic order in the massive (would-be partially massless) field. 

\subsection{Partial masslessness} 
\label{suse2:pm}

Let us begin by discussing the free theory of a partially massless field $\vp_{\mu\nu}$ in four dimensions~\cite{Deser:1983mm,Deser:2001pe,Deser:2001us,Deser:2001wx,Deser:2001xr,Zinoviev:2001dt,Deser:2004ji}. The latter is described by the Fierz-Pauli action~\cite{Fierz:1939ix} on a de Sitter background,
  \eq{
   I_{FP} = \int d^4 x \absq{\bar{g}} & \bigg \{ -  \frac{1}{2} \overline{\nabla}_{\rho}\vp_{\mu \nu} \overline{\nabla}^{\rho}\vp^{\mu \nu} + \frac{1}{2} \overline{\nabla}_{\rho}\vp \overline{\nabla}^{\rho}\vp -  \overline{\nabla}_{\rho}\vp \overline{\nabla}_{\s}\vp^{\rho \s} + \overline{\nabla}_{\rho}\vp_{\mu \nu} \overline{\nabla}^{\nu}\vp^{\mu \rho}  \\
   & + \ll \( \vp_{\mu\nu} \vp^{\mu\nu} -\frac{1}{2} \vp^2 \) - \frac{1}{2} m^2 \(\vp_{\mu\nu}\vp^{\mu\nu}  -   \vp^2 \) \bigg \}, \label{fierzpauli}
  }
where $\vp = \bar{g}^{\mu\nu}\vp_{\mu\nu}$, $\bar{g}_{\mu\nu}$ is the de Sitter metric, and $\ll$ is the cosmological constant. In eq.~\eqref{fierzpauli} the mass of the graviton satisfies the Higuchi bound~\cite{Higuchi:1986py}
  \eq{
  m^2 \ge \frac{2}{3} \ll,  \label{higuchi}
  }
which is the minimum value of the mass for which the action~\eqref{fierzpauli} is ghost free. 
  
While the Fierz-Pauli action admits a gauge symmetry when $m^2 \ra 0$, namely the standard diffeomorphism invariance of General Relativity, this action also features a gauge symmetry when $m^2$ saturates eq.~\eqref{higuchi}. This is the partially massless symmetry where $\vp_{\mu\nu}$ transforms as
  \eq{
  \d \vp_{\mu\nu} = \( \overline{\nabla}_{\mu} \overline{\nabla}_{\nu} + \frac{\ll}{3} \bar{g}_{\mu\nu} \) \xi(x). \label{pmsymmetry}
  }
In particular, the PM symmetry is responsible for removing the helicity-0 component of the otherwise massive graviton. Thus, in four dimensions partially massless fields propagate four degrees of freedom, in contrast to the five degrees of freedom of a massive graviton. 

In principle, a self-interacting theory of partially massless fields can be constructed order by order in the fields until, eventually, a pattern emerges which points towards a non-linear formulation. However, as discussed in the introduction, several no-go results suggest that such a theory does not exist as long as it contains at most two derivatives and one or several partially massless fields~\cite{Zinoviev:2006im,Deser:2013uy,deRham:2013wv,Garcia-Saenz:2014cwa,Garcia-Saenz:2015mqi}. Here we will consider the candidate partially massless theory of refs.~\cite{Hassan:2012gz,Hassan:2012rq,Hassan:2013pca,Hassan:2015tba} which can in principle circumvent these no-go results by adding a massless spin-2 field to the spectrum.


\subsection{Bimetric action} 
\label{suse2:allorders}

Consistent non-linear theories of two interacting spin-2 fields are described by the bimetric action~\cite{Hassan:2011zd,Hassan:2011ea}
  \eq{
  I[g,f] = m^2 \int d^4x \Big\{ \absq{g} \, R_g + \a^2 \absq{f} \, R_f -2 \mu^2 \absq{g} \sum_{n=0}^4 \b_n\, e_n(S) \Big\}, \label{se2:action}
  }
where $g_{\mu\nu}$ and $f_{\mu\nu}$ are two metrics whose perturbations correspond to linear combinations of a massless and a massive spin-2 fields. The constants $m$ and $\mu$ in eq.~\eqref{se2:action} have dimensions of energy while $\a$ and $\b_n$ are dimensionless constants. The functions $e_n(S)$ denote the elementary symmetric polynomials of the matrix $S$ which is defined by
  \eq{
  S^{\mu}{}_{\nu} = \(\sqrt{g^{-1}f}\)^{\mu}{}_{\nu}.
  }
In particular, we are interested in
  \eq{
  e_0(S) = 1, \qquad e_2(S) = \frac{1}{2} \Big [ \tr(S)^2 - \tr(S^2) \Big ], \qquad e_4(S) = \frac{\absq{f}}{\absq{g}}.
  }

The candidate non-linear theory of partially massless gravity that propagates an additional massless spin-2 field was identified in~\cite{Hassan:2012gz}. It corresponds to the bimetric action~\eqref{se2:action} with the following choice of $\b_n$ parameters 
  \eq{
  \b_0 = 3 \a^{-2} \b_2, \qquad \b_1 = 0, \qquad \b_3 = 0, \qquad \b_4 = 3 \a^2 \b_2, \label{se2:pmpoint}
  }
where $\b_2$ and $\a$ are left undetermined. Note that it is possible to recover the quadratic action of a partially massless field given in eq.~\eqref{fierzpauli} for several choices of $\b_n$ parameters. However, it is only for eq.~\eqref{se2:pmpoint} that the bimetric action admits a global version of the local PM symmetry, namely one where the function $\xi$ in eq.~\eqref{pmsymmetry} is constant. Furthermore, for this set of parameters the local PM symmetry can be extended up to sixth order in a derivative expansion of the equations of motion~\cite{Hassan:2013pca,Hassan:2015tba}. 

Interestingly, the $\b_n$ parameters given in eq.~\eqref{se2:pmpoint} make the potential of bimetric gravity symmetric upon exchange of the $g_{\mu\nu}$ and $f_{\mu\nu}$ metrics. They also guarantee that proportional solutions to the equations of motion, namely,
  \eq{
  g_{\mu\nu} = \bar{g}_{\mu\nu}, \qquad f_{\mu\nu} = c^2 \bar{g}_{\mu\nu},
  }
leave the constant $c$ undetermined. In particular, the equations of motion admit proportional de Sitter solutions where the cosmological constant is given by
  \eq{
  \ll = \frac{3\mu^2}{\a^2} (1+\a^2 c^2) \b_2. \label{se2:cc}
  }
This allows us to express the action entirely in terms of $\a$ and the dimensionful constants $m$, $\mu$, and $\ll$. 


\subsection{Bimetric action at quadratic and cubic order} 
\label{suse2:quadraticandcubic}

Let us now consider the bimetric action to quadratic and cubic order in the massive field $\vp_{\mu\nu}$. We begin by expanding the $g_{\mu\nu}$ and $f_{\mu\nu}$ metrics around a de Sitter background $\bar{g}_{\mu\nu}$ with a cosmological constant given by eq.~\eqref{se2:cc}\footnote{While it is common to refer to $\d A$ as the perturbation of the field $A$, for convenience we parametrize the perturbation of the metric $f_{\mu\nu}$ by $c^2 \d f_{\mu\nu}$ instead.},
  \eq{
  g_{\mu\nu} &= \bar{g}_{\mu\nu} + \d g_{\mu\nu}, \qquad f_{\mu\nu} = c^2 \( \bar{g}_{\mu\nu} + \d f_{\mu\nu} \). \label{se2:backgroundexpansion}
  }
In particular, we have
  \eq{
  g^{\mu\nu} &= \bar{g}^{\mu\nu} - \d g^{\mu\nu} + \d g^{\mu \a} \d g_{\a}^{\ph{\a}\nu} + \dots, \qquad f^{\mu\nu} = \frac{1}{c^2} \( \bar{g}^{\mu\nu} - \d f^{\mu\nu} + \d f^{\mu \a} \d f_{\a}^{\ph{\a}\nu} + \dots \), 
  }
where all indices are raised with the background metric $\bar{g}_{\mu\nu}$. The massless $h_{\mu\nu}$ and massive $\vp_{\mu\nu}$ fields correspond to linear combinations of $\d g_{\mu\nu}$ and $\d f_{\mu\nu}$. Up to normalization they are given by~\cite{Hassan:2012gz}
  \eq{
  \d g_{\mu\nu} = h_{\mu\nu} - \a^2 \, \vp_{\mu\nu}, \qquad \d f_{\mu\nu} = h_{\mu\nu} + \frac{1}{c^2}\, \vp_{\mu\nu}. \label{se2:fields}
  }
The factors of $\a^2$ and $c^2$ are important. For example, if we set $\a^2 c^2=1$ the $g_{\mu\nu} \leftrightarrow f_{\mu\nu}$ exchange symmetry of the potential extends also to the full action and leads to a vanishing cubic order action for the $\vp_{\mu\nu}$ field.\footnote{In fact, all the terms in the action containing an odd number of $\vp_{\mu\nu}$ fields vanish.}

The action of bimetric gravity~\eqref{se2:action} can then be written as
  \eq{
  I[\tilde{g},\vp] = \(1 + \a^2 c^2 \) m^2 \int d^4x \absq{\tilde{g}} \,\Big \{ R - 2\ll + {\textstyle \sum_{n=2}^{\infty}} \,\L_n \Big \},\label{se2:action2}
  }
where all contractions, covariant derivatives, and curvature tensors are defined with respect to the metric
  \eq{
  \tilde{g}_{\mu\nu} = \bar{g}_{\mu\nu} + h_{\mu\nu}, \qquad \tilde{g}^{\mu\nu} = \bar{g}^{\mu\nu} - h^{\mu\nu} + h^{\mu\a} h_{\a}^{\phantom{\a}\nu} + \dots \label{se2:masslesspert}
  }
In other words, in eq.~\eqref{se2:masslesspert} we have resummed the perturbations of the massless spin-2 fields to all orders. Hence, the bimetric action reduces to the action of a symmetric rank-2 tensor $\vp_{\mu\nu}$ coupled non-minimally to the metric $\tilde{g}_{\mu\nu}$. This is equivalent to expanding the action~\eqref{se2:action} around the off-shell metric $\tilde{g}_{\mu\nu}$ via
  \eq{
  g_{\mu\nu} = \tilde{g}_{\mu\nu} - \a^2 \vp_{\mu\nu}, \qquad f_{\mu\nu} = c^2 \tilde{g}_{\mu\nu} + \vp_{\mu\nu}. \label{se2:fields2}
  }
A relative factor of $c^2$ is still allowed since, in the absence of the $\vp_{\mu\nu}$ perturbations, the equations of motion for the $g_{\mu\nu}$ and $f_{\mu\nu}$ metrics both reduce to the Einstein equation, namely~$R_{\mu\nu} - \frac{1}{2} \tilde{g}_{\mu\nu} R + \ll \tilde{g}_{\mu\nu} = 0$. 
  
The Lagrangian density $\L_n$ in eq.~\eqref{se2:action2} depends on an $n$-number of $\vp_{\mu\nu}$ fields and contains single powers of the Ricci tensor. The fact that $\L_1$, which would be linear in $\vp_{\mu\nu}$, is missing from the action can be traced back to the relative sign and the extra factors of $\a^2$ and $c^2$ accompanying $\vp_{\mu\nu}$ in eqs.~\eqref{se2:fields} and~\eqref{se2:fields2}. Indeed, while the coefficients of $\vp_{\mu\nu}$ in the expansion around the de Sitter background, cf.~eqs.~\eqref{se2:backgroundexpansion} and~\eqref{se2:fields}, diagonalize the action at quadratic order, they also guarantee the absence of linear instabilities in the action when expanding around the off-shell metric $\tilde{g}_{\mu\nu}$ in eq.~\eqref{se2:fields2}.

At quadratic order in $\vp_{\mu\nu}$ we find that, up to total derivatives,
  \eq{
    \L_2 =  \frac{\a^2}{2 c^2} \bigg\{&  -  \frac{1}{2} \nabla_{\rho}\varphi_{\mu \nu} \nabla^{\rho}\varphi^{\mu \nu} + \frac{1}{2} \nabla_{\rho}\varphi \nabla^{\rho}\varphi -  \nabla_{\rho}\varphi \nabla_{\s}\varphi^{\rho \s} + \nabla_{\rho}\varphi_{\mu \nu} \nabla^{\nu}\varphi^{\mu \rho} + \frac{2\ll}{3} \vp_{\mu\nu} \vp^{\mu\nu}  \notag \\ 
    & -  \frac{\ll}{6} \vp^2 + G^{\mu \nu} \Big (- \frac{1}{2} \tilde{g}_{\mu \nu} \varphi_{\rho \s} \varphi^{\rho \s} + \frac{1}{4} \tilde{g}_{\mu \nu} \varphi^2 + 2 \varphi^{\rho}{}_{\nu} \varphi_{\mu \rho} -  \varphi  \varphi_{\mu \nu} \Big) \bigg\}, \label{se2:quadratic}
  }
where $\vp = \vp^{\mu}{}_{\mu}$ and $G_{\mu\nu} = R_{\mu \nu} -\tfrac{1}{2} R \tilde{g}_{\mu\nu} + \ll \tilde{g}_{\mu\nu}$ is the Einstein tensor with the cosmological constant term. In particular, if we expand $\L_2$ around the de Sitter background, cf.~eq.~\eqref{se2:masslesspert}, we recover the Fiertz-Pauli action given in eq.~\eqref{fierzpauli} where the mass of the spin-2 field $\vp_{\mu\nu}$ saturates the Higuchi bound~\cite{Higuchi:1986py},
  \eq{
  m^2_{\mathrm{PM}} = \frac{2}{3} \ll.
  }
Hence, at quadratic order the bimetric action describes a massless and a partially massless field propagating a total of $2+4$ degrees of freedom~\cite{Deser:1983mm,Deser:2001pe,Deser:2001us,Deser:2001wx,Deser:2001xr,Zinoviev:2001dt,Deser:2004ji}. Let us note that this action, which is defined to \emph{all orders} in the massless excitation $h_{\mu\nu}$, was also considered in the bottom-up, i.e.~order by order, approach to partially massless bimetric gravity of ref.~\cite{Joung:2014aba}.

At cubic order the Lagrangian density derived from the bimetric action~\eqref{se2:action} reads
  \eq{
  \L_3 = \l_3 \Big\{ & -  \frac{1}{2} \varphi^{\rho \s} \nabla_{\rho}\varphi^{\gamma \lambda} \nabla_{\s}\varphi_{\gamma \lambda} + \frac{1}{2} \varphi^{\rho \s} \nabla_{\rho}\varphi \nabla_{\s}\varphi -  \varphi^{\rho \s} \nabla_{\s}\varphi \nabla_{\gamma}\varphi_{\rho}{}^{\gamma} -  \varphi^{\rho \s} \nabla_{\s}\varphi_{\rho}{}^{\gamma} \nabla_{\gamma}\varphi  \notag \\
  &  + \varphi^{\rho \s} \nabla_{\gamma}\varphi \nabla^{\gamma}\varphi_{\rho \s} -  \frac{1}{4} \varphi \nabla_{\gamma}\varphi \nabla^{\gamma}\varphi -  \varphi^{\rho \s} \nabla^{\gamma}\varphi_{\rho \s} \nabla_{\lambda}\varphi_{\gamma}{}^{\lambda} + \frac{1}{2} \varphi \nabla^{\gamma}\varphi \nabla_{\lambda}\varphi_{\gamma}{}^{\lambda} \notag \\
  &+ 2 \varphi^{\rho \s} \nabla_{\s}\varphi_{\gamma \lambda} \nabla^{\lambda}\varphi_{\rho}{}^{\gamma} + \varphi^{\rho \s} \nabla_{\gamma}\varphi_{\s \lambda} \nabla^{\lambda}\varphi_{\rho}{}^{\gamma} -  \varphi^{\rho \s} \nabla_{\lambda}\varphi_{\s \gamma} \nabla^{\lambda}\varphi_{\rho}{}^{\gamma} -  \frac{1}{2} \varphi \nabla_{\gamma}\varphi_{\s \lambda} \nabla^{\lambda}\varphi^{\s \gamma} \notag \\
  & + \frac{1}{4} \varphi \nabla_{\lambda}\varphi_{\s \gamma} \nabla^{\lambda}\varphi^{\s \gamma} + \frac{1}{4}  \( R^{\rho \s} - \frac{1}{6} R \tilde{g}^{\rho\s} \) \Big ( 8 \varphi_{\rho}{}^{\gamma} \varphi_{\s}{}^{\lambda} \varphi_{\gamma \lambda}   -  2 \varphi_{\rho \s} \varphi_{\gamma \lambda} \varphi^{\gamma \lambda}  \notag \\
  &  - 4 \varphi_{\rho}{}^{\gamma} \varphi_{\s \gamma} \varphi  +  \varphi_{\rho \s} \varphi^2 \Big ) + \frac{1}{12} \Lambda  \Big ( 4 \varphi_{\rho}{}^{\gamma} \varphi^{\rho \s} \varphi_{\s \gamma}  -  \varphi  \varphi_{\s \gamma} \varphi^{\s \gamma} \Big ) \Big\}, \label{se2:cubic}
  }
where $\l_3$ is given by
  \eq{
  \l_3 = \frac{\a^2}{2c^4}(\a^2 c^2 - 1).
  }
This Lagrangian is different from the cubic order Lagrangian considered in~\cite{Joung:2014aba} which, in particular, does not contain any curvature tensors and is the simplest covariantization of the cubic order Lagrangian found in~\cite{Zinoviev:2006im}. Nevertheless it is possible to recover the cubic Lagrangian of~\cite{Joung:2014aba} via a field redefinition of the metric and the massive field. These field redefinitions are given by
  \eq{
  & \vp_{\mu\nu} \ra \vp_{\mu\nu} - \frac{(\a^2 c^2 - 1)}{4c^2} \,\vp_{\mu\rho}\vp^{\rho}{} _{\nu}, \label{se2:fieldred1} \\
  & \tilde{g}_{\mu\nu} \ra \tilde{g}_{\mu\nu} - \frac{\a^2  (\a^2 c^2 - 1) }{8 c^4} \Big ( \frac{5}{3} \tilde{g}_{\mu \nu } \varphi_{\alpha }{}^{\gamma } \varphi^{\alpha \beta } \varphi_{\beta \gamma } - \frac{3}{2} \tilde{g}_{\mu \nu } \varphi  \varphi_{\beta \gamma } \varphi^{\beta \gamma } + \frac{1}{3} \tilde{g}_{\mu \nu } \varphi^3 \notag \\
& \hspace{4.5cm}+ \varphi_{\alpha \beta } \varphi^{\alpha \beta } \varphi_{\mu \nu } - \varphi^2 \varphi_{\mu \nu } + 3 \varphi \varphi_{\mu }{}^{\alpha } \varphi_{\nu \alpha } - 4 \varphi_{\alpha \beta } \varphi_{\mu }{}^{\alpha } \varphi_{\nu }{}^{\beta } \Big). \label{se2:fieldred2}
  }
  %


\section{Gauge symmetries to cubic order in the action} 
\label{se3:pmsymmetry}

We now proceed to determine the PM gauge symmetry of the action order by order in the massive field $\vp_{\mu\nu}$. It will be convenient to find the transformation of the fields in a de Sitter background first and then to extend these results to all orders in the massless field $h_{\mu\nu}$. We will use a notation similar to that of~\cite{Joung:2014aba} where, schematically,
  \eq{
  I^{(n)} \sim \int d^4x \absq{\tilde{g}} \,\vp^{(n-2)}\nabla\vp\nabla\vp, \qquad \d^{(n)}_{\g} \O \sim \vp^{(n)} \nabla \nabla \g,  \label{se3:notation}
  }
for any field $\O$. In eq.~\eqref{se3:notation} $\g$ is a function of the coordinates that parametrizes the PM transformation and $I^{(0)}$ corresponds to the Einstein-Hilbert action.


\subsection{Zeroth order}

At zeroth order in the massive field the variation of the action reads
  \eq{
  \d I^{(0)} = \int d^4x \absq{\tilde{g}} \,\frac{\d I^{(0)}}{\d\tilde{g}^{\mu\nu}} \, \d^{(0)}\tilde{g}^{\mu\nu},
  }
which admits the standard gauge symmetry associated with diffeomorphisms, i.e.
  \eq{
  \d^{(0)} \tilde{g}_{\mu\nu} = \nabla_{(\mu} \xi_{\nu)},
  }
where we (anti)symmetrize indices with unit weight, e.g.~$\nabla_{(\mu}\xi_{\nu)} = \tfrac{1}{2} (\nabla_{\mu}\xi_{\nu} + \nabla_{\nu}\xi_{\mu})$. In particular, the non-linear analysis of ref.~\cite{Hassan:2015tba} suggests that, under PM transformations of $\vp_{\mu\nu}$, the metric transforms up to a convenient normalization by
  \eq{
  \d_{\g}^{(0)}\tilde{g}_{\mu\nu} = - \(\frac{\a^2c^2-1}{2 c^2}\) \nabla_{\mu}\nabla_{\nu} \g. \label{se3:zeroG}
  }
Since eq.~\eqref{se3:zeroG} looks just like a diffeomorphism, at this order in $\vp_{\mu\nu}$ we can always counter the transformation of the metric by a change of coordinates $\d x^{\mu} = \nabla^{\mu} \g$.  
  

\subsection{Second order}

At second order in $\vp_{\mu\nu}$, variation of the action yields
  \eq{
  \d I^{(1)} = \int d^4x \absq{\tilde{g}} \lB \frac{\d I^{(2)}}{\d \vp_{\mu\nu}} \, \d^{(0)} \vp_{\mu\nu} + \frac{\d I^{(0)}}{\d \tilde{g}^{\mu\nu}} \, \d^{(1)} \tilde{g}^{\mu\nu} \rB.
  }
The second term is the contribution from the Einstein-Hilbert action which vanishes in the de Sitter (dS) background $\bar{g}_{\mu\nu}$. Then $\d_{\g} I^{(1)}$ vanishes in the background if
  \eq{
  \d_{\g} I^{(1)} \big |_{dS} = 0  \qquad \Longrightarrow \qquad \d_{\g}^{(0)} \vp_{\mu\nu} \big |_{dS} = \(\overline{\nabla}_{\mu} \overline{\nabla}_{\nu} + \frac{\ll}{3} \bar{g}_{\mu\nu} \) \g, \label{se3:zeroPM}
  }
where $\overline{\nabla}_{\mu}$ denotes the covariant derivative with respect to $\bar{g}_{\mu\nu}$. Eq.~\eqref{se3:zeroPM} is the standard PM gauge transformation~\eqref{pmsymmetry} that, at quadratic order, is responsible for removing one of the degrees of freedom of what is otherwise a massive graviton~\cite{Deser:1983mm,Deser:2001pe,Deser:2001us,Deser:2001wx,Deser:2001xr,Zinoviev:2001dt,Deser:2004ji}.

We can extend the invariance of the action away from the de Sitter background provided that the metric transforms non-trivially under the partially massless gauge symmetry. Indeed, we have
  \eq{
  \d_{\g} I^{(1)} = 0  \qquad \Longrightarrow \qquad  & \d_{\g}^{(0)} \vp_{\mu\nu} = \( \nabla_{\mu} \nabla_{\nu} + \frac{\ll}{3} \tilde{g}_{\mu\nu} \) \g, \\
    & \d_{\g}^{(1)} \tilde{g}_{\mu\nu} = -\frac{\a^2}{2c^2} \big ( 2\nabla_{(\mu} \vp_{\nu)\rho} -\nabla_{\rho} \vp_{\mu\nu} \big ) \nabla^{\rho} \g. \label{se3:linearG}
  }
Note that this result was previously obtained in~\cite{Joung:2014aba} since their quadratic action agrees with ours up to normalization. Note also that the transformation of the metric is ambiguous up to total derivatives. The latter correspond to field-dependent diffeomorphisms, i.e.~to transformations of the form $\d^{(1)}\tilde{g}_{\mu\nu} = 2 \nabla_{(\mu} \xi_{\nu)}$ where $\xi^{\mu}$ depends linearly on $\vp_{\mu\nu}$. For example $\xi^{\mu}$ may be given by $\xi^{\mu} = \vp^{\mu\nu} \nabla_{\nu} \g$. These extra transformations do not change our results, however.


\subsection{Third order}

At third order in the massive field the variation of the action reads
  \eq{
  \d I^{(2)} = \int d^4x \absq{\tilde{g}} \lB \frac{\d I^{(3)}}{\d \vp_{\mu\nu}} \, \d^{(0)} \vp_{\mu\nu} + \frac{\d I^{(2)}}{\d \vp_{\mu\nu}} \, \d^{(1)} \vp_{\mu\nu} + \frac{\d I^{(2)}}{\d \tilde{g}^{\mu\nu}} \, \d^{(0)} \tilde{g}^{\mu\nu} + \frac{\d I^{(0)}}{\d \tilde{g}^{\mu\nu}} \, \d^{(2)} \tilde{g}^{\mu\nu} \rB. \label{se3:thirdOrder}
  }
Since the last term in eq.~\eqref{se3:thirdOrder} vanishes in the de Sitter background, vanishing of $\d_{\g} I^{(2)}$ determines the linear transformation of the massive field $\d^{(1)}_{\g} \vp_{\mu\nu}$. We find,  
  \eq{
  \d_{\g} I^{(2)} \big |_{dS} = 0  \quad \Longrightarrow \quad  & \d_{\g}^{(1)} \vp_{\mu\nu} \big |_{dS} = \( \frac{\a^2 c^2 -1}{2c^2} \) \( \overline{\nabla}_{(\mu} \vp_{\nu)\rho} \overline{\nabla}^{\rho} \g - \frac{1}{2} \overline{\nabla}_{\rho}\vp_{\mu\nu} \overline{\nabla}^{\rho} \g - \frac{\ll}{3} \vp_{\mu\nu} \,\g  \), \label{se3:linearPM}
  }
up to a transformation of the form $\d^{(0)}_{\tilde{\g}}\vp_{\mu\nu}$ where $\tilde{\g} = \vp \g$. The latter corresponds to a field-dependent PM transformation which can be trivially cancelled via an appropriate shift of $\d^{(2)}_{\g}\vp_{\mu\nu}$. Although these kind of transformations do affect the higher order transformations of the fields they leave our results unchanged. The normalization used in eq.~\eqref{se3:zeroG} was chosen so as to simplify the expression for $\d_{\g}^{(1)}\vp_{\mu\nu}$ given above.  
  
We now have all the ingredients necessary to carry out the analysis of global symmetries in the bimetric theory. We should note that while $\d_{\g}^{(1)} \tilde{g}_{\mu\nu}$ agrees with ref.~\cite{Joung:2014aba}, $\d_{\g}^{(1)} \vp_{\mu\nu}$ does not. The differences in $\d_{\g}^{(1)} \vp_{\mu\nu}$ can be traced back to (1) the non-linear field redefinition given in eq.~\eqref{se2:fieldred1}; and (2) an extra diffeomorphism that results from having a non-zero $\d_{\g}^{(0)}\tilde{g}_{\mu\nu}$ transformation. However, it is not immediately clear whether these extra ingredients of the bimetric theory are enough to change the no-go results found in~\cite{Joung:2014aba}. 

Before we turn to the analysis of global symmetries let us generalize the above results away from the de Sitter background. Vanishing of $\d_{\g} I^{(2)}$ under PM transformations yields
  \eq{
  \d_{\g} I^{(2)} = 0  \quad \Longrightarrow \quad  & \d_{\g}^{(1)} \vp_{\mu\nu} = \( \frac{\a^2 c^2 -1}{2c^2} \) \(\nabla_{(\mu} \vp_{\nu)\rho} \nabla^{\rho} \g - \frac{1}{2} \nabla_{\rho}\vp_{\mu\nu} \nabla^{\rho} \g - \frac{\ll}{3} \vp_{\mu\nu} \,\g  \), \\
  & \d_{\g}^{(2)} \tilde{g}_{\mu\nu} = \frac{\a^2}{c^2} \(\frac{\a^2 c^2 -1}{2c^2}\) \( \nabla_{(\mu} \vp_{\nu) \s} - \frac{1}{2} \nabla_{\s} \vp_{\mu\nu} \) \vp^{\s\rho} \nabla_{\rho} \g,
  }
up to trivial transformations similar to the ones discussed above and in the previous section.


\section{Global symmetries}
\label{se:global}

Let us now consider the global symmetries of the candidate PM theory. As in any gauge theory with non-trivial boundary conditions, not all of the generators of gauge symmetries are proportional to the constraints. Indeed, some generators receive boundary contributions that lead to finite, non-vanishing charges. For asymptotically de Sitter spacetimes the set of diffeomorphisms leads to an $SO(1,4)$ global symmetry group. Since the partially massless theory possesses an enlarged set of gauge symmetries it is natural to expect an enhancement of the global symmetry group, although this is not automatically guaranteed.\footnote{See~\cite{Hinterbichler:2015nua} for the construction of charges in the quadratic partially massless theory.} 

This question was recently considered in~\cite{Joung:2014aba} in an order-by-order approach to partially massless bimetric gravity. The authors of~\cite{Joung:2014aba} first compute the algebra generated by the PM and diff transformations to lowest order in the fields. This field-independent algebra is the algebra of global symmetries provided that there exist gauge parameters that leave the background invariant. To lowest order in the fields the global symmetry algebra of the partially massless theory is enhanced from $SO(1,4)$ to $SO(1,5)$~\cite{Joung:2014aba}. Since in de Sitter space there is no analog of the Coleman-Mandula theorem, this enhancement of spacetime symmetries may be consistent at higher orders, i.e.~it may be realized in an interacting theory.

Let us first check that the bimetric theory admits an enhanced set of global symmetries. The reason why this check is non-trivial is that the commutator of symmetries depends on the transformation of $\d_{\g}^{(1)} \vp_{\mu\nu}$ which differs from that considered in~\cite{Joung:2014aba}. It is convenient to write down the PM transformations of the massless and massive fields up to linear order in the fields. First we expand around the de Sitter background using eq.~\eqref{se2:masslesspert}. Then from eqs.~\eqref{se3:zeroG}, \eqref{se3:zeroPM}, and \eqref{se3:linearPM} we have
  \eq{
  \d_{\g} h_{\mu\nu} = & \l_1 \bnabla_{\mu} \bnabla_{\nu} \g - \frac{\l_1}{2} \( 2 \bnabla_{(\mu} h_{\nu)\s} - \bnabla_{\s}h_{\mu\nu} \) \partial^{\s} \g - \l_2 \( 2 \bnabla_{(\mu} \vp_{\nu) \s} - \bnabla_{\s} \vp_{\mu\nu} \) \partial^{\s}\g,\\
  \d_{\g} \vp_{\mu\nu} = & \(\overline{\nabla}_{\mu} \overline{\nabla}_{\nu} + \frac{\ll}{3} \bar{g}_{\mu\nu} \) \g -\frac{1}{2} \( 2\bnabla_{(\mu} h_{\nu) \s} - \bnabla_{\s} h_{\mu\nu} \) \partial^{\s}\g + \frac{\ll}{3} h_{\mu\nu} \g + \l_1 \frac{\ll}{3} \vp_{\mu\nu} \g \notag \\
  & - \frac{\l_1}{2} \( 2\bnabla_{(\mu} \vp_{\nu) \s} - \bnabla_{\s} \vp_{\mu\nu} \) \p^{\s} \g ,
  }
where $\l_1$ and $\l_2$ are given by
  \eq{
  \l_1 = - \frac{\a^2 c^2 -1}{2 c^2}, \qquad \l_2 = \frac{\a^2}{2c^2}.
  }
On the other hand, the transformations of $h_{\mu\nu}$ and $\vp_{\mu\nu}$ under diffeomorphisms, denoted here by $\tilde{\d}_{\xi}$, are given by the standard expressions
  \eq{
  \tilde{\d}_{\xi} h_{\mu\nu} =& 2\bnabla_{(\mu} \xi_{\nu)} + \xi^{\s} \bnabla_{\s} h_{\mu\nu} + 2 \bnabla_{(\mu} \xi^{\s} h_{\nu)\s}, \qquad \tilde{\d}_{\xi} \vp_{\mu\nu} = \xi^{\s} \bnabla_{\s} \vp_{\mu\nu} + 2 \bnabla_{(\mu} \xi^{\s} \vp_{\nu)\s}. \label{se3:diff}
  }

To zeroth order in the fields the algebra of PM transformations ($\d_{\g}$) and diffeomorphisms ($\tilde{\d}_{\xi}$) closes,
  \eq{
  [\tilde{\d}_{\xi}, \tilde{\d}_{\eta}] \O = &\tilde{\d}_{\chi} \O, \qquad [\d_{\g}, \tilde{\d}_{\eta}] \O = \d_{\tau} \O, \qquad [\d_{\g},\d_{\b}] \O = \tilde{\d}_{\t} \O, \label{se3:commutators}
  }
where $\O$ denotes either the massless $h_{\mu\nu}$ or massive $\vp_{\mu\nu}$ fields, and the parameters $\chi^{\mu}$, $\tau$, and $\theta^{\mu}$ are given by
  \eq{
 \chi^{\mu} \p_{\mu} &\equiv [\eta, \xi] = \( \eta^{\rho} \bnabla_{\s} \xi^{\mu} - \xi^{\s} \bnabla_{\s} \eta^{\mu} \) \p_{\mu}, \label{se3:comm1}\\
  \tau & \equiv [\eta,\g] = \eta^{\s} \p_{\s} \g, \\
 \t^{\mu} \p_{\mu} & \equiv [\b, \g] = \frac{ (\a^2 c^2 + 1)^2}{16 c^4} \( \p^{\s}\g \bnabla_{\s} \p^{\mu} \b -\p^{\s} \b \bnabla_{\s} \p^{\mu} \g \) \p_{\mu}. \label{se3:comm3}
  }
These results agree with~\cite{Joung:2014aba} except that the commutator $[\d_{\g},\tilde{\d}_{\eta}] h_{\mu\nu}$ does not vanish. This reflects the fact that the metric has a non-zero PM transformation at lowest order and, more importantly, that it leads to a consistent algebra. 

In order to determine the global symmetry algebra one must first find the parameters $\g$ and $\xi$ which leave the background $\tilde{g}_{\mu\nu} = \bar{g}_{\mu\nu}$ and $\vp_{\mu\nu} = 0$ invariant. We have two cases.
  \begin{description}
  
  \item{\bf Case 1:} diffeomorphisms and PM transformations vanish independently, whereby the only solution to
  \eq{
  0 = \d_{\g} \bar{g}_{\mu\nu} = \l_1 \bnabla_{\mu} \bnabla_{\nu} \g, \qquad 0 = \d_{\g} \vp_{\mu\nu} = \(\overline{\nabla}_{\mu} \overline{\nabla}_{\nu} + \frac{\ll}{3} \bar{g}_{\mu\nu} \) \g,
  }
  is $\g = 0$. In this case the global symmetry algebra is the standard $SO(1,4)$ symmetry algebra of de Sitter space obtained from eq.~\eqref{se3:comm1} and generated by the solutions to
  \eq{
  0 = \d_{\xi} \bar{g}_{\mu\nu} = 2 \bnabla_{(\mu} \xi_{\nu)}.
  }

  \item{\bf Case 2:} use a diffeomorphism to undo the PM transformation of the background. In this case we can define a new PM transformation $\d'_{\g} = \d_{\g} - \tilde{\d}_{\zeta}$ with $\zeta^{\mu} = -\tfrac{\l_1}{2} \p^{\mu} \g$ where, to lowest order in the fields, $\d'_{\g} g_{\mu\nu} = 0$ while the transformation of $\vp_{\mu\nu}$ is left unchanged. Then the background is left invariant provided that
  \eq{
  0 = \d_{\xi} \bar{g}_{\mu\nu} = 2 \bnabla_{(\mu} \xi_{\nu)}, \qquad 0 = \d'_{\g} \vp_{\mu\nu} = \(\overline{\nabla}_{\mu} \overline{\nabla}_{\nu} + \frac{\ll}{3} \bar{g}_{\mu\nu} \) \g. \label{se3:killinglike}
  }
  These equations do admit non-trivial solutions. Furthermore, since the algebra~\eqref{se3:commutators} is left unchanged by the prescription $\d_{\g} \ra \d'_{\g}$, the solutions to eq.~\eqref{se3:killinglike} lead to an $SO(1,5)$ algebra as shown explicitly in~\cite{Joung:2014aba}.

  \end{description}

This is not the end of the story, however, since the closure of the algebra~\eqref{se3:commutators} may not hold to higher orders. This would render the global symmetry algebra inconsistent at higher orders. To check whether this is the case for the candidate PM bimetric theory, let us write down the transformations of the fields under the $\d'_{\g}$ PM transformations. Up to linear order in the fields these are given by,
  \eq{
  \d'_{\g} h_{\mu\nu} = & - \l_2 \( 2 \bnabla_{(\mu} \vp_{\nu) \s} - \bnabla_{\s} \vp_{\mu\nu} \) \partial^{\s}\g,\\
  \d'_{\g} \vp_{\mu\nu} = & \(\overline{\nabla}_{\mu} \overline{\nabla}_{\nu} + \frac{\ll}{3} \bar{g}_{\mu\nu} \) \g -\frac{1}{2} \( 2\bnabla_{(\mu} h_{\nu) \s} - \bnabla_{\s} h_{\mu\nu} \) \partial^{\s}\g + \frac{\ll}{3} h_{\mu\nu} \g + \l_1 \frac{\ll}{3} \vp_{\mu\nu} \g \notag \\
  & - \frac{\l_1}{2} \( 2\bnabla_{(\mu} \vp_{\nu) \s} - \bnabla_{\s} \vp_{\mu\nu} \) \p^{\s} \g + \frac{\l_1}{2} \( \p^{\s} \g \bnabla_{\s} \vp_{\mu\nu} + 2 \bnabla_{(\mu} \p^{\s} \g \vp_{\nu)\s} \).
  }
The crucial point is that on the parameters generating the global symmetries, i.e.~on the solutions to eq.~\eqref{se3:killinglike}, these transformations reduce to the transformations considered in~\cite{Joung:2014aba} for appropriate values of the coefficients $\l_1$ and $\l_2$. In particular, this implies that at linear order in the massive field, the commutator of two PM transformations does not close~\cite{Joung:2014aba}
  \eq{
  [\d'_{\g}, \d'_{\b}] \vp_{\mu\nu} = \tilde{\d}_{\xi} \vp_{\mu\nu} + \frac{(\a^2c^2+1)^2}{16c^4} C_{\mu\nu}, \label{se3:nonclosure}
  }
where $\tilde{\d}_{\xi} \vp_{\mu\nu}$ is given by eq.~\eqref{se3:diff} for some $\xi^{\mu}$ while $C_{\mu\nu}$ is a function of $\vp_{\mu\nu}$ and the parameters $\g$ and $\b$. Since the commutator of two PM transformations should close into a diffeomorphism, cf.~\eqref{se3:commutators}, the second term in the RHS of eq.~\eqref{se3:nonclosure} should vanish.

Thus the algebra of global symmetries becomes inconsistent at higher orders. One should note that this analysis is not valid for the choice $\a^2 c^2 =1$, in which case both the left and right-hand sides of eq.~\eqref{se3:nonclosure} vanish. Also note that the action contains only factors of $\a^2$ and $c^2$, so setting $\a^2 c^2 = -1$ is valid insofar as it leads to a real action. However, this choice of parameters leads to an action~\eqref{se2:action2} where our calculations can no longer be trusted. That the expansion of the action breaks down for this choice of parameters can be seen directly from eq.~\eqref{se2:fields} where the choice $\a^2 c^2 = -1$ degenerates and leads to an inconsistent parametrization of the massless and massive fields. Indeed, when $\a^2 c^2 = -1$ the quadratic action can no longer be diagonalized in terms of massless and massive excitations. Interestingly, this is precisely the set of parameters for which the equations of motion of the bimetric theory reproduce those of conformal gravity at lowest order in derivatives~\cite{Hassan:2013pca,Hassan:2015tba}.

Since the bimetric theory exists to all orders and can be rendered free of pathologies~\cite{Hassan:2016abc}, we conclude that the global symmetry algebra $SO(1,5)$ is accidental. Thus, the candidate PM theory admits only the standard $SO(1,4)$ global symmetries of de Sitter space.


\section{Gauge symmetries to fourth order}
\label{se:fourthorder}

Let us return to the gauge symmetries of the candidate partially massless bimetric theory. The absence of an enhanced global symmetry group may be taken as a hint of a larger problem. Indeed, in the analysis of~\cite{Joung:2014aba} the non-closure of the global symmetry algebra shows that the local PM symmetry cannot be extended beyond cubic order. However, the analysis of~\cite{Joung:2014aba} is not valid for for the choice $\a^2 c^2 =1$ of the bimetric theory where terms odd in the PM field vanish. Therefore, a natural question to ask is whether the PM gauge symmetry can be extended beyond cubic order in the bimetric theory for any choice of parameters. In principle this is a hopeless endeavor since success at fourth order does not guarantee success at fifth or higher orders. A more promising approach would be to search for an additional first class constraint in the Hamiltonian formulation of bimetric gravity that would be responsible for removing the helicity-0 mode from the massive spin-2 field. However, the square root structure in the potential in eq.~\eqref{se2:action} makes this a complicated task. Furthermore, the fact that an order-by-order approach fails in massive gravity already at fourth order makes a similar calculation in the bimetric setup worthwhile. 

At fourth order in $\vp_{\mu\nu}$ the variation of the action reads
  \eqsp{
  \d I^{(3)} = \int d^4x \absq{\tilde{g}} \bigg \{ \frac{\d I^{(4)}}{\d \vp_{\mu\nu}} \, \d^{(0)} \vp_{\mu\nu} + \frac{\d I^{(3)}}{\d \vp_{\mu\nu}} \, \d^{(1)} \vp_{\mu\nu}  +& \frac{\d I^{(2)}}{\d \vp_{\mu\nu}} \, \d^{(2)} \vp_{\mu\nu} + \frac{\d I^{(3)}}{\d \tilde{g}^{\mu\nu}} \, \d^{(0)} \tilde{g}^{\mu\nu} \\
  +& \frac{\d I^{(2)}}{\d \tilde{g}^{\mu\nu}} \, \d^{(1)} \tilde{g}^{\mu\nu} + \frac{\d I^{(0)}}{\d \tilde{g}^{\mu\nu}} \, \d^{(3)} \tilde{g}^{\mu\nu}  \bigg \}, \label{se4:fourthOrderVar}
  }
where $I^{(4)}$ is the fourth-order action described by the Lagrangian density~\eqref{ap1:fourthorder} given in Appendix~\ref{ap:fourthorder}. As before let us first consider variation of the action in the de Sitter background $\bar{g}_{\mu\nu}$. Then the last term in eq.~\eqref{se4:fourthOrderVar} vanishes. On the other hand, the variations $\d^{(0)}_{\g} \tilde{g}_{\mu\nu}$ and $\d^{(1)}_{\g} \tilde{g}_{\mu\nu}$ of the massless field \emph{do} contribute to the variation of the action. These have already been determined in eqs.~\eqref{se3:zeroG} and \eqref{se3:linearG}, and the only undetermined term in eq.~\eqref{se4:fourthOrderVar} is the second order variation of $\vp_{\mu\nu}$, namely $\d^{(2)}_{\g} \vp_{\mu\nu}$. 

Unfortunately, the presence of $\d^{(n)}g_{\mu\nu}$ terms is not sufficient to overcome the no-go result found in massive gravity~\cite{deRham:2013wv}. There, it was found terms of the form $\d_{\g} I^{(3)} \sim \int d^4x \vp \overline{\nabla} \vp \overline{\nabla} \vp \g$ cannot be cancelled for any choice of $\d^{(2)}_{\g} \vp_{\mu\nu}$. Indeed, in the bimetric theory we find that, for a relatively simple transformation of the form
  \eqsp{
  \d^{(2)}_{\g} \vp_{\mu\nu} =  &\( \frac{1+\a^4 c^4}{2c^4} \) \( \overline{\nabla}_{(\mu} \vp_{\nu) \s} -\frac{1}{2} \overline{\nabla}_{\s} \vp_{\mu\nu} \) \vp^{\s\rho} \p_{\rho} \g \\
  &+ \frac{\ll}{2} \(\frac{1 + \a^2 c^2 }{2 c^2}\)^2 \( \frac{1}{3} \bar{g}_{\mu\nu} \vp_{\rho\s} \vp^{\rho\s} - \vp_{\mu}{}^{\a} \vp_{\a\nu} \)\g, \label{se4:quadraticPM}
  }
variation of the action yields
  \eqst{
  \d_{\g} I^{(3)} = & \frac{\a^2 m^2(1+\a^2 c^2)^3 \ll}{48 c^6} \int d^4x \absq{\bar{g}} \,\g\bigg \{ 2 \varphi^{\rho \s} \overline{\nabla}_{\rho}\varphi \overline{\nabla}_{\s}\varphi -2 \varphi^{\rho \s} \overline{\nabla}_{\rho}\varphi^{\gamma \lambda} \overline{\nabla}_{\s}\varphi_{\gamma \lambda}   \\
  & - 2 \varphi^{\rho \s} \overline{\nabla}_{\s}\varphi \overline{\nabla}_{\gamma}\varphi_{\rho}{}^{\gamma} + 2 \varphi^{\rho \s} \overline{\nabla}_{\gamma}\varphi \overline{\nabla}^{\gamma}\varphi_{\rho \s} -  \varphi \overline{\nabla}_{\gamma}\varphi \overline{\nabla}^{\gamma}\varphi -  \varphi \overline{\nabla}_{\s}\varphi^{\s \gamma} \overline{\nabla}_{\lambda}\varphi_{\gamma}{}^{\lambda}  \\
  & - 2 \varphi^{\rho \s} \overline{\nabla}^{\gamma}\varphi_{\rho \s} \overline{\nabla}_{\lambda}\varphi_{\gamma}{}^{\lambda} + 2 \varphi \overline{\nabla}^{\gamma}\varphi \overline{\nabla}_{\lambda}\varphi_{\gamma}{}^{\lambda} + 2 \varphi_{\rho}{}^{\gamma} \varphi^{\rho \s} \overline{\nabla}_{\lambda}\overline{\nabla}_{\gamma}\varphi_{\s}{}^{\lambda} - 2 \varphi_{\rho}{}^{\gamma} \varphi^{\rho \s} \overline{\nabla}_{\lambda}\overline{\nabla}^{\lambda}\varphi_{\s \gamma}  \\
  & + 4 \varphi^{\rho \s} \overline{\nabla}_{\s}\varphi_{\gamma \lambda} \overline{\nabla}^{\lambda}\varphi_{\rho}{}^{\gamma} - 2 \varphi^{\rho \s} \overline{\nabla}_{\lambda}\varphi_{\s \gamma} \overline{\nabla}^{\lambda}\varphi_{\rho}{}^{\gamma} -  \varphi \overline{\nabla}_{\gamma}\varphi_{\s \lambda} \overline{\nabla}^{\lambda}\varphi^{\s \gamma} + \varphi \overline{\nabla}_{\lambda}\varphi_{\s \gamma} \overline{\nabla}^{\lambda}\varphi^{\s \gamma} \bigg \}.
  }
It is possible to further simplify the variation of the action at the cost of introducing more terms in $\d^{(2)}_{\g} \vp_{\mu\nu}$. For example, adding the following terms to eq.~\eqref{se4:quadraticPM}
  \eqst{
  \d^{(2)}_{\g} \vp_{\mu\nu} \ra \,\, & \d^{(2)}_{\g} \vp_{\mu\nu} + \(\frac{1 + \a^2 c^2 }{4 c^2}\)^2 \bigg \{ -\frac{2}{3} \ll \bar{g}_{\mu\nu} \vp_{\rho\s} \vp^{\rho\s} \g + 2 \ll \vp_{\mu}{}^{\a} \vp_{\a\nu} \g - \overline{\nabla}_{\s} \( \vp^{\s}{}_{\rho} \overline{\nabla}_{\mu} \vp_{\nu}{}^{\rho} \g \) \\
   & + \overline{\nabla}_{\mu} \( \vp_{\nu}{}^{\s} \overline{\nabla}_{\s} \vp^{\rho}{}_{\rho} \g \)  - \overline{\nabla}_{\s} \( \vp_{\rho\mu} \overline{\nabla}_{\nu} \vp^{\s\rho} \g \)  - \overline{\nabla}_{\s} \( \vp^{\s\rho} \vp_{\rho\mu} \overline{\nabla}_{\nu} \g \) + \frac{1}{2} \overline{\nabla}_{\mu} \( \overline{\nabla}_{\s} \vp^{\s}{}_{\nu} \vp \g \) \\
   & + \overline{\nabla}_{\mu} \( \vp_{\s\rho} \overline{\nabla}_{\nu} \vp^{\s\rho} \g \)  -\frac{1}{2} \vp \overline{\nabla}_{\mu} \vp \overline{\nabla}_{\nu} \g - \frac{1}{2} \overline{\nabla}_{\mu} \vp \overline{\nabla}_{\nu} \vp \g + \frac{1}{2} \vp \overline{\nabla}_{\mu}\overline{\nabla}_{\nu} \vp \g  +(\mu \leftrightarrow \nu ) \bigg\},
  }
reduces the variation of the action to
  \eqsp{
  \d_{\g} I^{(3)} = & \frac{\a^2 m^2(1+\a^2 c^2)^3 \ll}{48 c^6} \int  d^4x \absq{\bar{g}} \,\g \bigg \{ 4 \varphi^{\rho \s} \overline{\nabla}_{\s}\varphi_{\gamma \lambda} \overline{\nabla}^{\lambda}\varphi_{\rho}{}^{\gamma} -2 \varphi^{\rho \s} \overline{\nabla}_{\rho}\varphi^{\gamma \lambda} \overline{\nabla}_{\s}\varphi_{\gamma \lambda}  \\
  & \hspace{2.5cm} - 2 \varphi^{\rho \s} \overline{\nabla}_{\lambda}\varphi_{\s \gamma} \overline{\nabla}^{\lambda}\varphi_{\rho}{}^{\gamma} -  \varphi \overline{\nabla}_{\gamma}\varphi_{\s \lambda} \overline{\nabla}^{\lambda}\varphi^{\s \gamma} + \varphi \overline{\nabla}_{\lambda}\varphi_{\s \gamma} \overline{\nabla}^{\lambda}\varphi^{\s \gamma} \bigg \}.
  \label{se4:minvar}
  }
One would expect that adding more terms to $\d^{(2)}_{\g} \vp_{\mu\nu}$ could render the action invariant under PM transformations, however unnatural the transformation may become. Unfortunately there are no such terms and there is no $\d^{(2)}_{\g} \vp_{\mu\nu}$ transformation that leads to an invariant action, in agreement with the general results of~\cite{Joung:2014aba}. Thus the partially massless symmetry cannot be extended beyond cubic order and the bimetric theory with the $\b_n$ parameters given in eq.~\eqref{se2:pmpoint} propagates a total of $2+5$ degrees of freedom corresponding to a massless and a massive spin-2 field.


\subsection{Generalizing the quartic action}

It is now reasonable to ask whether there exists any action for which the PM symmetry can be extended beyond the cubic order. In order to investigate this, let us once again consider equation \eqref{se4:fourthOrderVar}, where we now let $I^{(4)}$ be the most general fourth-order two-derivative action, rather than the one given in eq.~\eqref{ap1:fourthorder}. In the de Sitter background $\bar{g}_{\mu \nu}$, this action can be schematically written as
\eq{
	I^{(4)} = \int d^4x \absq{\bar{g}} \left\{ \vp^2 \overline{\nabla} \vp\overline{\nabla} \vp + \Lambda \vp^4 \right\},
}
with all possible index contractions and arbitrary coefficients in front of all terms. The second order variation of $\vp_{\mu \nu}$, $\d_\g^{(2)} \vp_{\mu \nu}$, remains undetermined in eq.~\eqref{se4:fourthOrderVar}, whereas all other terms have already been fixed at lower orders. We then consider the most general $\d_\g^{(2)} \vp_{\mu \nu}$ transformation with at most two derivatives. It is schematically given by
  \eq{
  \d_\g^{(2)} \vp = \overline{\nabla}\vp\overline{\nabla} \vp \g + \vp \overline{\nabla}^2 \vp \g + \vp \overline{\nabla} \vp \overline{\nabla} \g + \vp^2 \overline{\nabla}^2 \g + \ll \vp^2 \g.
  }

Up to cubic order in the the massive field the variation of the action takes the following form
\eqsp{
	\d I^{(3)} = \int d^4x \absq{\bar{g}} \g \bigg \{& \vp^2 \overline{\nabla}^4 \vp + \vp \overline{\nabla} \vp \overline{\nabla}^3 \vp + \vp \overline{\nabla}^2 \vp \overline{\nabla}^2 \vp + \overline{\nabla} \vp \overline{\nabla} \vp \overline{\nabla}^2 \vp \\
	& + \Lambda \vp^2 \overline{\nabla}^2 \vp + \Lambda \vp \overline{\nabla} \vp \overline{\nabla} \vp + \Lambda^2 \vp^3  \bigg \}, \label{se5:fourthOrderVar2}
}
where once again all index contractions and coefficients have been omitted. The question is now if it is possible to choose the parameters of $I^{(4)}$ and $\d_\g^{(2)} \vp_{\mu \nu}$ in such a way that this variation vanishes and the partially massless symmetry is preserved up to quartic order. It turns out that one can chose parameters such that all terms in eq.~\eqref{se5:fourthOrderVar2} vanish, except for terms of the form $\Lambda \vp \overline{\nabla} \vp \overline{\nabla} \vp \,\g$. In fact, the best one can do is reduce the variation of the action given in eq.~\eqref{se5:fourthOrderVar2} to that of the bimetric theory given in eq.~\eqref{se4:minvar}. The quartic action for which this maximal cancellation of terms takes place is precisely that of bimetric gravity.\footnote{Note that the quartic action can be changed from that of bimetric gravity at the cost of adding extra terms to the transformation $\d_\g^{(2)} \vp_{\mu \nu}$. We have ignored such terms throughout this paper since they do not remove the non-trivial contributions to eq.~\eqref{se4:fourthOrderVar} that originate from lower order terms.} Hence the bimetric theory is the closest we can get to a working PM theory with only two spin-2 fields.


\section{Conclusions}
\label{se:conclusions}

In this paper we have analyzed the gauge and global symmetries of the candidate partially massless bimetric gravity up to fourth order in the massive field. We have seen that the action of the theory reduces to the action of a massive spin-2 field coupled non-minimally to gravity. Using the appropriate parametrization of the massless and massive fields, cf.~eqs.\eqref{se2:fields} and~\eqref{se2:fields2}, we have shown that the cubic order action does not vanish. In particular, this action reduces to the cubic action of a partially massless field studied in~\cite{Zinoviev:2006im}, as well as the covariantization considered in~\cite{Joung:2014aba}, after suitable field redefinitions. Crucially, we have seen that the global symmetry analysis of~\cite{Joung:2014aba} extends to the bimetric setup. This implies that the $SO(1,5)$ global symmetry of the candidate PM bimetric theory is accidental and only the standard $SO(1,4)$ symmetry of de Sitter space survives non-linearly.

The absence of an $SO(1,5)$ global symmetry is not surprising in light of our second result. Namely, that the PM gauge symmetry cannot be extended beyond cubic order in the action, in agreement with the results presented in~\cite{Joung:2014aba}. In fact, there is no quartic action which is invariant under an extension of the PM symmetry. Thus the presence of an additional massless spin-2 field is not sufficient to render the quartic interactions of a massive spin-2 field invariant under PM transformations. Our results fall in line with similar results found in the literature that rule out the existence of a non-linear theory of partially massless gravity describing one or a multiplet of PM fields~\cite{Zinoviev:2006im,Deser:2013uy,deRham:2013wv,Garcia-Saenz:2014cwa,Garcia-Saenz:2015mqi}.

One way to avoid these no-go results is to further enlarge the spectrum of the theory, i.e.~to add lower or higher spin fields that transform non-trivially under the PM gauge symmetry. For example, refs.~\cite{Gwak:2015vfb,Gwak:2015jdo} consider a three-dimensional model of colored gravity interacting non-trivially with $SU(N)$ vector fields. Upon spontaneous breaking of the SU(N) symmetry it is seen that all except one of the spin-two fields become partially massless. However, this construction works only in three dimensions and it is not obvious whether the partially massless symmetry can be extended beyond quadratic order. A related four-dimensional theory where the partially massless symmetry can be extended to all orders will be presented in~\cite{Apolo:2016ort}.


\section*{Acknowledgments}

It is a pleasure to thank Nico Wintergerst for several helpful discussions and Massimo Taronna for valuable correspondence. This work was supported by a grant from the Swedish Research Council. L.A. also thanks Bo Sundborg for support. The perturbative expansion of the action and the analysis of gauge symmetries were performed using the Mathematica tensor package \emph{xAct}~\cite{xact,Nutma:2013zea}. 


\appendix

\section{Fourth order action}
\label{ap:fourthorder}

The quartic Lagrangian in the action~\eqref{se2:action2} is complicated but not particularly illuminating. Its distinguishing feature is that terms without derivatives, i.e.~terms proportional to the cosmological constant, depend on the parameters $\a^2$ and $c^2$, unlike the quadratic~\eqref{se2:quadratic} and cubic~\eqref{se2:cubic} Lagrangian densities. Up to total derivatives the quartic Lagrangian density reads
  \eq{
  \L_4 = \l_4 \bigg \{ &\frac{\Lambda}{6(1 -  \a^2 c^2 + \a^4 c^4)} \bigg [ 4(5 - 2 \a^2 c^2 + 5 \a^4 c^4)\, \varphi_{\rho}{}^{\gamma} \varphi^{\rho \s} \varphi_{\s}{}^{\lambda} \varphi_{\gamma \lambda} - 4(1 + \a^2 c^2)^2 \, \varphi^{\rho}{}_{\rho} \varphi_{\s}{}^{\lambda} \varphi^{\s \gamma} \varphi_{\gamma \lambda}  \notag \\ 
  & -  (1 + 6 \a^2 c^2 + \a^4 c^4) \, \varphi_{\rho \s} \varphi^{\rho \s} \varphi_{\gamma \lambda} \varphi^{\gamma \lambda} + 8 \a^2 c^2 \, \varphi^{\rho}{}_{\rho} \varphi^{\s}{}_{\s} \varphi_{\gamma \lambda} \varphi^{\gamma \lambda} -  \a^2 c^2 \, \varphi^{\rho}{}_{\rho} \varphi^{\s}{}_{\s} \varphi^{\gamma}{}_{\gamma} \varphi^{\lambda}{}_{\lambda} \bigg ] \notag \\
  & + \frac{1}{3} \( R^{\rho \s} -\frac{1}{8} R g^{\rho\s} \) \Big ( 96 \varphi_{\rho}{}^{\gamma} \varphi_{\s}{}^{\lambda} \varphi_{\gamma}{}^{\mu} \varphi_{\lambda \mu} - 16 \varphi_{\rho \s} \varphi_{\gamma}{}^{\mu} \varphi^{\gamma \lambda} \varphi_{\lambda \mu} - 24  \varphi_{\rho}{}^{\gamma} \varphi_{\s \gamma} \varphi_{\lambda \mu} \varphi^{\lambda \mu} \notag \\
  & + 12  \varphi_{\rho \s} \varphi^{\gamma}{}_{\gamma} \varphi_{\lambda \mu} \varphi^{\lambda \mu} - 48 \varphi_{\rho}{}^{\gamma} \varphi_{\s}{}^{\lambda} \varphi_{\gamma \lambda} \varphi^{\mu}{}_{\mu} + 12 \varphi_{\rho}{}^{\gamma} \varphi_{\s \gamma} \varphi^{\lambda}{}_{\lambda} \varphi^{\mu}{}_{\mu} -  2 \varphi_{\rho \s} \varphi^{\gamma}{}_{\gamma} \varphi^{\lambda}{}_{\lambda} \varphi^{\mu}{}_{\mu}  \Big )   \notag \\
  & +16 \varphi^{\rho \s} \varphi^{\gamma \lambda} \nabla_{\s}\varphi_{\lambda \mu} \nabla_{\gamma}\varphi_{\rho}{}^{\mu} - 8 \varphi_{\rho}{}^{\gamma} \varphi^{\rho \s} \nabla_{\s}\varphi^{\lambda \mu} \nabla_{\gamma}\varphi_{\lambda \mu} + 4 \varphi^{\rho}{}_{\rho} \varphi^{\s \gamma} \nabla_{\s}\varphi^{\lambda \mu} \nabla_{\gamma}\varphi_{\lambda \mu} \notag \\ 
  & + 8 \varphi_{\rho}{}^{\gamma} \varphi^{\rho \s} \nabla_{\s}\varphi^{\lambda}{}_{\lambda} \nabla_{\gamma}\varphi^{\mu}{}_{\mu} - 4 \varphi^{\rho}{}_{\rho} \varphi^{\s \gamma} \nabla_{\s}\varphi^{\lambda}{}_{\lambda} \nabla_{\gamma}\varphi^{\mu}{}_{\mu} - 16 \varphi_{\rho}{}^{\gamma} \varphi^{\rho \s} \nabla_{\gamma}\varphi^{\mu}{}_{\mu} \nabla_{\lambda}\varphi_{\s}{}^{\lambda}  \notag \\ 
  & + 8 \varphi^{\rho}{}_{\rho} \varphi^{\s \gamma} \nabla_{\gamma}\varphi^{\mu}{}_{\mu} \nabla_{\lambda}\varphi_{\s}{}^{\lambda} - 16 \varphi^{\rho \s} \varphi^{\gamma \lambda} \nabla_{\gamma}\varphi_{\rho}{}^{\mu} \nabla_{\lambda}\varphi_{\s \mu} - 16 \varphi^{\rho \s} \varphi^{\gamma \lambda} \nabla_{\s}\varphi_{\rho \gamma} \nabla_{\lambda}\varphi^{\mu}{}_{\mu}  \notag \\ 
  & + 16 \varphi^{\rho \s} \varphi^{\gamma \lambda} \nabla_{\gamma}\varphi_{\rho \s} \nabla_{\lambda}\varphi^{\mu}{}_{\mu} - 16 \varphi_{\rho}{}^{\gamma} \varphi^{\rho \s} \nabla_{\gamma}\varphi_{\s}{}^{\lambda} \nabla_{\lambda}\varphi^{\mu}{}_{\mu} + 8 \varphi^{\rho}{}_{\rho} \varphi^{\s \gamma} \nabla_{\gamma}\varphi_{\s}{}^{\lambda} \nabla_{\lambda}\varphi^{\mu}{}_{\mu}  \notag \\ 
  & + 16 \varphi_{\rho}{}^{\gamma} \varphi^{\rho \s} \nabla_{\lambda}\varphi^{\mu}{}_{\mu} \nabla^{\lambda}\varphi_{\s \gamma} - 8 \varphi^{\rho}{}_{\rho} \varphi^{\s \gamma} \nabla_{\lambda}\varphi^{\mu}{}_{\mu} \nabla^{\lambda}\varphi_{\s \gamma} - 2 \varphi_{\rho \s} \varphi^{\rho \s} \nabla_{\lambda}\varphi^{\mu}{}_{\mu} \nabla^{\lambda}\varphi^{\gamma}{}_{\gamma}  \notag \\ 
  & + \varphi^{\rho}{}_{\rho} \varphi^{\s}{}_{\s} \nabla_{\lambda}\varphi^{\mu}{}_{\mu} \nabla^{\lambda}\varphi^{\gamma}{}_{\gamma} - 16 \varphi^{\rho \s} \varphi^{\gamma \lambda} \nabla_{\gamma}\varphi_{\rho \s} \nabla_{\mu}\varphi_{\lambda}{}^{\mu} - 16 \varphi_{\rho}{}^{\gamma} \varphi^{\rho \s} \nabla^{\lambda}\varphi_{\s \gamma} \nabla_{\mu}\varphi_{\lambda}{}^{\mu}  \notag \\ 
  & + 8 \varphi^{\rho}{}_{\rho} \varphi^{\s \gamma} \nabla^{\lambda}\varphi_{\s \gamma} \nabla_{\mu}\varphi_{\lambda}{}^{\mu} + 4 \varphi_{\rho \s} \varphi^{\rho \s} \nabla^{\lambda}\varphi^{\gamma}{}_{\gamma} \nabla_{\mu}\varphi_{\lambda}{}^{\mu} - 2 \varphi^{\rho}{}_{\rho} \varphi^{\s}{}_{\s} \nabla^{\lambda}\varphi^{\gamma}{}_{\gamma} \nabla_{\mu}\varphi_{\lambda}{}^{\mu} \notag \\ 
  & - 16 \varphi^{\rho \s} \varphi^{\gamma \lambda} \nabla_{\lambda}\varphi_{\gamma \mu} \nabla^{\mu}\varphi_{\rho \s} + 8 \varphi^{\rho \s} \varphi^{\gamma \lambda} \nabla_{\mu}\varphi_{\gamma \lambda} \nabla^{\mu}\varphi_{\rho \s} + 32 \varphi^{\rho \s} \varphi^{\gamma \lambda} \nabla_{\lambda}\varphi_{\s \mu} \nabla^{\mu}\varphi_{\rho \gamma}  \notag \\ 
  & - 8 \varphi^{\rho \s} \varphi^{\gamma \lambda} \nabla_{\mu}\varphi_{\s \lambda} \nabla^{\mu}\varphi_{\rho \gamma} + 32 \varphi_{\rho}{}^{\gamma} \varphi^{\rho \s} \nabla_{\gamma}\varphi_{\lambda \mu} \nabla^{\mu}\varphi_{\s}{}^{\lambda} - 16 \varphi^{\rho}{}_{\rho} \varphi^{\s \gamma} \nabla_{\gamma}\varphi_{\lambda \mu} \nabla^{\mu}\varphi_{\s}{}^{\lambda} \notag \\ 
  &  + 16 \varphi_{\rho}{}^{\gamma} \varphi^{\rho \s} \nabla_{\lambda}\varphi_{\gamma \mu} \nabla^{\mu}\varphi_{\s}{}^{\lambda} - 8 \varphi^{\rho}{}_{\rho} \varphi^{\s \gamma} \nabla_{\lambda}\varphi_{\gamma \mu} \nabla^{\mu}\varphi_{\s}{}^{\lambda} - 16 \varphi_{\rho}{}^{\gamma} \varphi^{\rho \s} \nabla_{\mu}\varphi_{\gamma \lambda} \nabla^{\mu}\varphi_{\s}{}^{\lambda}  \notag \\ 
  & + 8 \varphi^{\rho}{}_{\rho} \varphi^{\s \gamma} \nabla_{\mu}\varphi_{\gamma \lambda} \nabla^{\mu}\varphi_{\s}{}^{\lambda} - 4 \varphi_{\rho \s} \varphi^{\rho \s} \nabla_{\lambda}\varphi_{\gamma \mu} \nabla^{\mu}\varphi^{\gamma \lambda} + 2 \varphi^{\rho}{}_{\rho} \varphi^{\s}{}_{\s} \nabla_{\lambda}\varphi_{\gamma \mu} \nabla^{\mu}\varphi^{\gamma \lambda}  \notag \\ 
  & + 2 \varphi_{\rho \s} \varphi^{\rho \s} \nabla_{\mu}\varphi_{\gamma \lambda} \nabla^{\mu}\varphi^{\gamma \lambda} -  \varphi^{\rho}{}_{\rho} \varphi^{\s}{}_{\s} \nabla_{\mu}\varphi_{\gamma \lambda} \nabla^{\mu}\varphi^{\gamma \lambda} \bigg \}, \label{ap1:fourthorder}
  }
where $\l_4 = \frac{\a^2}{32c^6} (1 - \a^2 c^2 + \a^4 c^4)$.


\ifprstyle
	\bibliographystyle{apsrev4-1}
\else
	\bibliographystyle{utphys2}
\fi

\bibliography{pmfg}


\end{document}